# Fluctuation theorems and 1/*f* noise from a simple matrix

Ralph V. Chamberlin,[1] Sumiyoshi Abe,[2,3] Bryce F. Davis,[1] Priscilla E. Greenwood,[4] Andrew S.H. Shevchuk[1]

[1]*Department of Physics, Arizona State University, Tempe, AZ 85287-1504, USA*
[2]*Department of Physical Engineering, Mie University, Mie 514-8507, Japan*
[3]*Institute of Physics, Kazan Federal University, Kazan 420008, Russia*
[4]*Department of Mathematics, University of British Columbia, Vancouver, BC V6T 1Z2, Canada*

**Abstract:** Here we present a model for a small system combined with an explicit entropy bath that is comparably small. The dynamics of the model is defined by a simple matrix, M. Each row of M corresponds to a macrostate of the system, e.g. net alignment, while the elements in the row represent microstates. The constant number of elements in each row ensures constant entropy, which allows reversible fluctuations, similar to information theory where a constant number of bits allows reversible computations. Many elements in M come from the microstates of the system, but many others come from the bath. Bypassing the bath states yields fluctuations that exhibit standard white noise; whereas with bath states the power spectral density varies as $S(f) \propto 1/f$ over a wide range of frequencies, *f*. Thus, the explicit entropy bath is the mechanism of 1/*f* noise in this model. Both forms of the model match Crooks' fluctuation theorem exactly, indicating that the theorem applies not only to infinite reservoirs, but also to finite-sized baths. The model is used to analyze measurements of 1/*f*-like noise from a sub-micron tunnel junction.

Fluctuation theorems [1-5] provide fundamental formulas that have been used to describe the thermal properties of small systems that may be far from equilibrium. These formulas have been applied to the out-of-equilibrium behavior of several small systems including: stretching of RNA molecules [6], information-to-energy conversion [7-9], and particles driven by an external field [10,11]. Of course the theorems should also apply to fluctuations of small systems around equilibrium, but what if the thermal bath is similarly small? Nanothermodynamics was originally developed to describe the thermal properties of a large ensemble of small systems [12,13], which has been extended to treat the dynamics of individual small systems with their own local bath [14,15]. Here we describe a model based on a single matrix M that contains a maximum number of system states $\Omega_0$, with a comparable number of bath states. We show that if the bath states are bypassed, the matrix model yields a power spectral density that exhibits standard white noise and agrees with one value from the Crooks fluctuation theorem [1]. However if the explicit bath states are included, the matrix model yields $1/f$-like noise and agrees with a different value from the Crooks theorem. Finally, the matrix model is used to analyze measurements of voltage noise from a sub-micron-sized tunnel junction [16].

The matrix model is based on information theory [17], adapted to treat thermal fluctuations. The matrix model includes bath states that maintain maximum entropy during fluctuations of a binary system, similar to "garbage" states in information theory. Neighboring rows of M are connected by an $\Omega_0$-to-$\Omega_0$ map that yields reversible fluctuations, similar to the one-to-one map that yields reversible computations (e.g. Fig. 3a of Ref. [17]). Specific Hamiltonians that may accomplish the transfer of information are discussed in Refs. [17,18], but here we focus on the general principle that entropy must remain maximized if fluctuations are to be reversible. Experimental evidence establishing that information is a physical quantity comes from the minimum work necessary to erase a single bit [9].

All system states in the matrix model are degenerate, with an explicit local bath that is comparable to system size and depends on system entropy, thereby violating multiple assumptions required for thermal properties to be governed solely by Boltzmann's factor [19]. Thus, the resulting thermal fluctuations differ from those found from standard statistical mechanics and stochastic thermodynamics [20], which are based on Boltzmann's factor alone. Indeed, because all system states are degenerate, the thermal properties are governed entirely by entropy, similar to microcanonical behavior. And as expected, finite-sized fluctuations depend on the ensemble. To compare: Boltzmann's factor favors low *energy* states of the system when *energy* is transferred to an infinite heat reservoir; whereas the matrix model favors low *entropy* states of the system when *entropy* is transferred to the finite entropy bath. Non-degenerate states in the system can be accommodated by combining Boltzmann's factor from energy transfer with a nonlinear correction from entropy transfer [14,15], but here we emphasize that the entropy-transfer mechanism is the crucial ingredient for 1/*f*-like noise. Using Boltzmann's factor for energy transfer generally yields deviations from pure 1/*f* behavior [21].

Thermal fluctuations exhibiting 1/*f*-like noise were first reported in 1925 [22,23]. Similar 1/*f*-like noise has since been found in virtually all types of materials, as well as in electronic, magnetic, and quantum devices [24-26], and even in biological systems [27,28] and human preferences [29,30]. No single mechanism can explain all details in the dynamics of such diverse systems. Nevertheless, an Ising-like model has been found to simulate measured 1/*f*-like noise, including temperature-dependent slopes shown by various metal films, and tri-modal histograms exhibited by spin glasses and nanopore systems [21]. Here we study the basic cause of 1/*f* noise in the Ising-like model. The key ingredient is a nonlinear correction to Boltzmann's factor that can be justified in several ways [14,15,31,32]. One mechanism comes from conservation of energy by including Hill's subdivision potential to treat non-extensive contributions from finite-sized fluctuations. Another comes from the statistics of indistinguishable particles. Here we focus on a third mechanism, where maximum entropy is maintained by transferring entropy between the system and its bath.

Consider an isolated small system that fluctuates into a low-entropy state. Although seeming to violate the 2$^{nd}$ law of thermodynamics, at least three possible explanations have been proposed. Total entropy may decrease temporarily if the system is small enough [2,33]. The time-averaged Gibbs entropy should be used instead of the time-dependent Boltzmann expression [34,35]. Alternatively, because information theory [7,8] requires a transfer of entropy when knowledge about the system changes, a decrease in entropy of the system implies an increase in entropy elsewhere, consistent with measurements of bit erasure [9]. Information theory is also consistent with using Lagrange multipliers for higher moments that maximize the total entropy of a fluctuating system and its environment. Here we treat the transferred entropy exactly by including explicit bath states that slow down the dynamics of the system.

For simplicity, assume that the system contains an even number ($N$) of binary degrees of freedom ("spins"), so that each spin may be up (+1) or down (–1). If the alignment of the system is unknown, then the total multiplicity is always $\Omega_{All} = 2^N$. Whereas if the net alignment ($m$) is known, then the multiplicity of each macrostate is given by the binomial coefficient $\Omega_m = N!/\{[½(N+m)]![½(N–m)]!\}$. Using Boltzmann's definition, the alignment entropy of the system is $S_m = k_B \ln(\Omega_m)$. The maximum multiplicity occurs when spins are half-up and half-down: $\Omega_0 = N!/[(½N)!]^2$, yielding the maximum entropy $S_0 = k_B \ln(\Omega_0)$. Fluctuations reduce this alignment entropy, $S_m \leq S_0$. The matrix model is based on the assumption that if $S_m<S_0$ because $m \neq 0$, an amount of entropy $S_0 - S_m$ must have been transferred from the system to its bath. Thus, the total entropy of the system plus bath remains maximized at $S_0$, which never decreases.

The dynamics of the system is governed by a rectangular matrix, M, see Fig. 1. (Note that the standard transition matrix of the alignment process is easily obtained from M.) The elements of M are $M_{ij}$, with $N+1$ rows ($-N/2 \leq i \leq N/2$) from the net alignments of the system $m=2i$, and $\Omega_0$ columns ($1 \leq j \leq \Omega_0$) from all microstates of the system plus bath. The middle row of M contains $\Omega_0$ non-zero elements ($M_{0j}=\pm 1$), one for each configuration of the unaligned system, with no states for the bath. The sign of the element governs how the alignment of the system will change: $M_{0j}=+1$ if $m$ is to increase and $M_{0j}=-1$ if $m$ is to decrease. Thus, in the middle row there are $\Omega_0/2$ elements having $M_{0j}=+1$, and an equal number having $M_{0j}=-1$. Other rows have $\Omega_m<\Omega_0$ non-zero elements, from the number of system states for $m \neq 0$. The remaining $\Omega_0-\Omega_m$ elements in each row have $M_{ij}=0$, representing bath states. The system states in each row, $M_{ij}=+1$ or $-1$, have the appropriate ratio for the probability that the alignment will increase or decrease, respectively. Specifically, there are $\Omega_m[\frac{1}{2}(N-m)/N] = (N-1)!/\{[\frac{1}{2}(N+m)]![\frac{1}{2}(N-m)-1]!\}$ elements having $M_{ij} = +1$; and $\Omega_m[\frac{1}{2}(N+m)/N] = (N-1)!/\{[\frac{1}{2}(N+m)-1]![\frac{1}{2}(N-m)]!\}$ elements having $M_{ij} = -1$. Adding extra 0 elements, as might be expected for large baths in contact with small systems, lowers the frequency where the system starts $1/f$-like behavior, but does not alter the general features of the $1/f$ regime. Using Stirling's formula, the ratio of bath states to system states is: $\frac{2(N+1)}{\sqrt{2\pi N}} - 1$. Thus, for $N \geq 6$ there are more bath states than system states; but the square-root dependence keeps the ratio relatively small, so that even for $N=24$ there are only about 3 times as many bath states as system states. Because the number of bath states depends on system entropy, our model yields fluctuations that differ from most other treatments that are based solely on Boltzmann's factor that assumes an infinite heat reservoir.

The dynamics of the model involves a time step (d$t$) that has two parts. In part one, an element is chosen at random from the current row of M. In part two, this element is used to determine how the row might change: if $M_{ij} = 0$ there is no change $j \rightarrow j$; if $M_{ij} = +1$ the row increases by one $j \rightarrow j+1$; if $M_{ij} = -1$ the row decreases by one $j \rightarrow j-1$. In other words, all possible changes in alignment are determined by $m(t+dt)=m(t)+2M_{ij}$. The dynamics continues by repeating parts one and two. For standard Monte-Carlo simulations without explicit bath states, part one is modified by choosing only non-zero elements (states with $M_{ij} = 0$ are bypassed), so that $m(t)$ changes every step. To realistically simulate measurements over a wide dynamic range we use an averaging time $\tau$. Specifically, if $\tau=1$ $m(t)$ is recorded every step, if $\tau=10$ $m(t)$ is averaged over 10 steps before recording, etc. We use averaging times up to $\tau=10^6$ steps, with $2^{17}$ data points per simulation, yielding up to $1.31 \times 10^{11}$ steps per simulation. Smoother curves are obtained by simulating the system ~20 times using different initial conditions, but intrinsic noise is retained by analyzing each simulation separately before averaging. To compare systems of different size we use relative alignment, $\lambda(t)=m(t)/N$.

The solid (open) symbols in Fig. 2 come from histograms of $\lambda(t)$ with (without) bath states, from the matrix having $N=24$. Each point in the histogram gives the likelihood of the alignment. Thus, when $\tau=1$ (to avoid averaging between alignments) a logarithmic plot (Fig. 2) yields the entropy as a function of alignment. Without explicit bath states (open squares), the $\tau=1$ histogram matches the binomial distribution (dashed curve), as expected for non-interacting binary degrees of freedom with no local bath. For $\tau \gg 1$ this discrete binomial evolves to a continuous Gaussian, as expected from the central-limit theorem. The width of the Gaussian decreases with increasing $\tau$, and approaches a delta function for very large $\tau$, as expected from the law of large numbers. Indeed, after the system has had time to explore all states, $\tau \gg N$, a Gaussian fit [$\ln(\Omega) \sim -\lambda^2/2\sigma^2$] to the central part of the peak yields a variance of $\sigma^2 \approx 0.95/\tau$, so that $\sigma^2 \approx 9.5 \times 10^{-7}$ for $\tau = 10^6$. Such inverse-$\tau$ dependence is expected for long-time averaging of normal fluctuations. In contrast, simulations with explicit bath states yield constant entropy for $\tau = 1$ (horizontal dotted line), evolving into a broad Gaussian with excess wings for $\tau \gg 1$. Again fitting the central part of the peak to a Gaussian, the variance is $\sigma^2 \approx 7.7 \times 10^{-3}$ for $\tau = 10^6$, which extrapolates towards $\sigma^2 \sim 3.2 \times 10^{-3}$ as $\tau \to \infty$. However, because the range of alignments is limited ($-1 \leq \lambda \leq +1$), this central Gaussian must also eventually approach a delta function for $\tau \gg 2^N = 1.67 \times 10^7$.

Crooks' fluctuation theorem is consistent with both forms of behavior shown in Fig. 2. Furthermore, since both forms involve all states explicitly, Crooks' fluctuation theorem is a consequence of detailed balance. First consider the open squares for fluctuations of the system without explicit bath states. Note that because the states of the system are degenerate, entropy alone governs all thermal properties, so that fluctuations are identical in the microcanonical ensemble with no bath, and canonical ensemble with infinite heat reservoir. Let the "forward" step change the alignment from $m$ to $m+2$, and the "reverse" step change $m+2$ to $m$. From the fraction of non-zero elements in row $i = m/2$ having $M_{ij} = +1$, the forward step has probability $P_m(+2) = \frac{1}{2}(N-m)/N$, the reverse step has $P_{m+2}(-2) = \frac{1}{2}(N+m+2)/N$, and their ratio is $R_S = P_m(+2)/P_{m+2}(-2) = (N-m)/(N+m+2)$. The Crooks' fluctuation theorem states that this ratio should come from the difference in entropy: $R_S = \exp[(S_{m+2} - S_m)/k_B]$. Indeed, using the entropy of the system from the binomial coefficient yields $\exp[(S_{m+2} - S_m)/k_B] = (N-m)/(N+m+2)$, matching the ratio $R_S$ as expected from the fluctuation theorem. Here $R_S \neq 1$ implies irreversible work must be done to change the net alignment of the system [4].

Now consider fluctuations of the system plus explicit entropy bath. For large systems with $m \approx \pm N$ the dynamics is very slow because the system plus bath spends most of its time exploring bath states that do not change $m$. Specifically, the probability of a forward step is reduced by a factor $\Omega_m/\Omega_0=[(\frac{1}{2}N)!]^2/\{[\frac{1}{2}(N+m)]![\frac{1}{2}(N-m)]!\}$, with the reverse step similarly reduced $\Omega_{m+2}/\Omega_0= [(\frac{1}{2}N)!]^2/\{[\frac{1}{2}(N+m)+1]![\frac{1}{2}(N-m)-1]!\}$. Note that the ratio of these factors is $\Omega_m/\Omega_{m+2}=(N+m+2)/(N-m)$, so that the ratio of forward to reverse steps is simply $R_{S+B}=R_S \Omega_m/\Omega_{m+2}=1$. Here $R_{S+B}=1$ implies that the system may change its net alignment reversibly [4]. Thus, from Crooks' fluctuation theorem the total entropy of the system plus bath should be constant, as ensured by the constant number of elements in each row and shown by the solid squares in Fig.2.

The solid symbols in Fig. 3 show the power spectral density of the alignment process, $S(f)$. The data come from the same simulations used for Fig. 2, and from 5 other systems with sizes $N<24$. The relative alignment as a function of discrete time is converted to the power spectral density using a discrete Fourier transform: $S(f)=\left|\frac{1}{j}\sum_{t=0}^{j-1}\lambda(t)\exp(-2\pi i f t / j)\right|^2$. The spectra are smoothed by linear regression using a sliding frequency range; then spectra with different $\tau$ are matched using a weighted average [21]. The diagonal dashed line indicates exact $1/f$ noise. The single system with $N=24$ shows $1/f$-like behavior over at least 4 orders of magnitude in $f$, but the frequency range decreases with decreasing $N$. Open triangles show the lowest-frequency mode of each system, given by the single fully-aligned state divided by the total number of all states, $f_0 \propto 1/\Omega_0 = [(\frac{1}{2}N)!]^2/N!$, which includes $\Omega_0 - 1$ bath states. Without explicit bath states, the symbols near the bottom of Fig. 3 exhibit white noise over at least 8 orders of magnitude in $f$, with no 1/$f$ regime,

Figure 4 presents a useful way to focus on the 1/$f$-like behavior, where $S(f)$ is multiplied by frequency so that 1/$f$ noise becomes horizontal. Indeed, the horizontal dashed line is the same equation as the diagonal dashed line in Fig. 3 showing 1/$f$ noise. The solid lines showing 1/$f$-like behavior over 4 and 2 orders of magnitude in frequency come from the matrix model with $N$=24 and $N$=16, respectively. Triangles mark the lowest-frequency modes from the fully-aligned states $f_0 \propto 1/\Omega_0$ (as in Fig. 3). Diamonds mark the expected higher-frequency modes from progressively less-aligned states of the $N$=24 matrix: $f_1=f_0 N$, $f_2=f_0 N(N-1)/2$, $f_3=f_0 N(N-1)(N-2)/6$, etc.

The dotted line in Fig. 4 comes from the sum of responses from one $N$=24 matrix plus ten times the $N$=16 matrix. The symbols are from measurements [16] of voltage noise across a metal-insulating-metal (MIM) tunnel junction measured at constant currents of (A) 65 and (B) 105 µA. The MIM junction is small enough (0.3 µm on both sides) that deviations from pure 1/$f$ noise can be seen. The dominant features are three distinct maxima. Although qualitatively similar to the oscillations in the simulations, a quantitative analysis yields unrealistic values for the frequencies, as follows. From the published measurements we find characteristic frequencies of $f_{A0}$ =7.2 Hz, $f_{A1}$ = 585 Hz, $f_{A2}$ = 9.52 kHz and $f_{B0}$ =10.1 Hz, $f_{B1}$ = 865 Hz, $f_{B2}$ = 12.9 kHz. Averaging the ratio of the lowest two frequencies yields $f_1/f_0$ = 84±2, implying that the MIM junction would need $N$≈84 binary degrees of freedom if these peaks were to come from the two lowest frequencies of a single matrix. This $N$ yields $f_0 \sim [(\frac{1}{2}N)!]^2 / N! \sim 1.68 \times 10^{-24}$, which would require that the microscopic dynamics be an unphysical 24 orders of magnitude faster than the measured $f_0$. More likely the distinct maxima come from independent subsystems inside the sample, supporting the original interpretation of the measurements. Indeed, there is qualitative similarity between the experimental data and dotted line from the independent matrices, including the net slope when plotted as in Fig. 4 showing deviations from pure 1/*f* noise. Quantitative values can be obtained by assuming microscopic dynamics at a frequency of $f_{max}$ ~ $10^{10}$ Hz. Then, using $f_0 = f_{max}/\Omega_0$ for the characteristic frequency, the published data yield *N*~33, 26, and 22 for the effective number of binary degrees of freedom causing the three measured peaks. The matrix model is too simplistic to capture all details in these measured spectra, but the similarity suggests that it may be necessary to include information entropy to capture the essence of the 1/*f*-like behavior.

We have shown that a simple system with explicit bath states fluctuates differently than the isolated system, and also differently from the system coupled to an infinite reservoir. Without explicit bath states the fluctuations yield well-known white noise. With explicit bath states the fluctuations yield 1/*f*-like noise, which is measured universally at lower frequencies, signifying the final approach to thermal equilibrium. Having explicit bath states also maintains maximum entropy. Thus, at least if the fluctuations are slow enough, the 2$^{nd}$ law of thermodynamics may be a fundamental physical law, not just a statistical rule-of-thumb. Additional tests of the matrix model will come from measuring systems with small enough baths that they exhibit the return to white noise at ultra-low frequencies, as shown in Fig. 3. From Stirling's approximation of the multiplicities, the model predicts that the minimum frequency for 1/*f*-like noise should decrease exponentially with increasing number of degrees of freedom. At these ultra-low frequencies, the system has time to explore all states of the system and its bath, thereby achieving the true thermal equilibrium. Because it is usually difficult to calculate all contributions to entropy exactly, the matrix model provides a simplistic way to simulate how strict adherence to the 2$^{nd}$ law of thermodynamics yields a basic mechanism for 1/*f*-like noise.

We thank S. Deffner and G. H. Wolf for helpful comments. Most of the simulations utilized the A2C2 computing facility at Arizona State University. RVC, BFD, and ASHS are grateful for financial support from the ARO via W911NF-11-1-0419. SA was supported in part by JSPS and the Ministry of Education and Science of the Russian Federation (the program of competitive growth of Kazan Federal University).

**Figure Legends**

FIG. 1. System states (left) and rectangular matrix M (right) for an $N=4$ spin system. Non-zero elements in M correspond to system states, with the fraction of $M_{ij} = +1$ or $-1$ from the probability that inverting a spin at random will increase or decrease the alignment, respectively. Elements with $M_{ij} = 0$ are explicit bath states that pause the dynamics of the system.

FIG. 2 (color online). Histograms, converted to the logarithm of multiplicity, from the dynamics of the $N=24$ matrix. Solid (open) symbols are from simulations with (without) explicit bath states. Different shapes come from different averaging times, from $\tau=1$ to $10^6$ steps. The most-probable values in the wings come from the allowed alignments, with lower values from averaging between them.

FIG. 3 (color online). Power spectral density of noise as a function of frequency. Solid symbols are from simulations of systems with several sizes ($N$). Note that S($f$) is multiplied by $N$ to scale the power from normal fluctuations, and log($f$) is multiplied by 10 to match the dB scale. The wide frequency range is obtained by averaging the Fourier transform from several time-series sequences, with averaging times of $\tau=1$ to $10^6$ steps. The diagonal dashed line shows exact 1/$f$ behavior. The open triangles identify the frequency of the slowest response rate, from 10 log($f$) = 53.5 – 10 log($\Omega_0$). The hexagonal symbols are from the $N=24$ matrix, but with all bath states bypassed.

FIG. 4 (color online). Log-log plot of noise power spectral density multiplied by frequency. The dashed line (from the same equation as the dashed line in Fig. 3) shows exact 1/$f$ noise. Solid lines, from the $N$=24 and 16 matrices M with explicit bath states, show a $1/f$-like regime with distinct oscillations. Triangles mark the lowest frequency for transitions out of the fully-aligned state: $f_0 \propto 1/\Omega_0 = [(\tfrac{1}{2}N)!]^2/N!$. Diamonds mark progressively higher frequencies expected for the less-aligned states of the $N$=24 system. The dotted line comes from the response of one $N$=24 matrix added to ten times the $N$=16 matrix, with the amplitude offset for clarity. Symbols are from measurements of voltage noise across a submicron tunnel junction [16], with the amplitude and frequency offset for clarity.

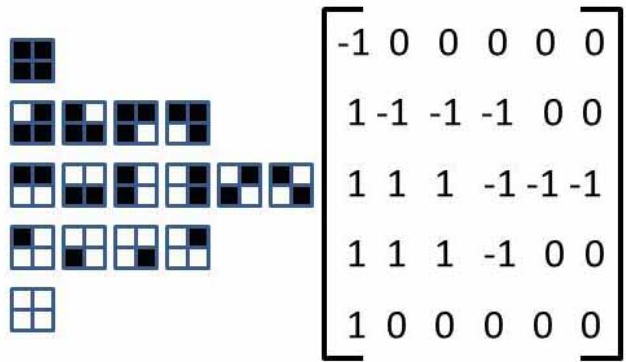

**Fig. 1**

**Fig. 2**

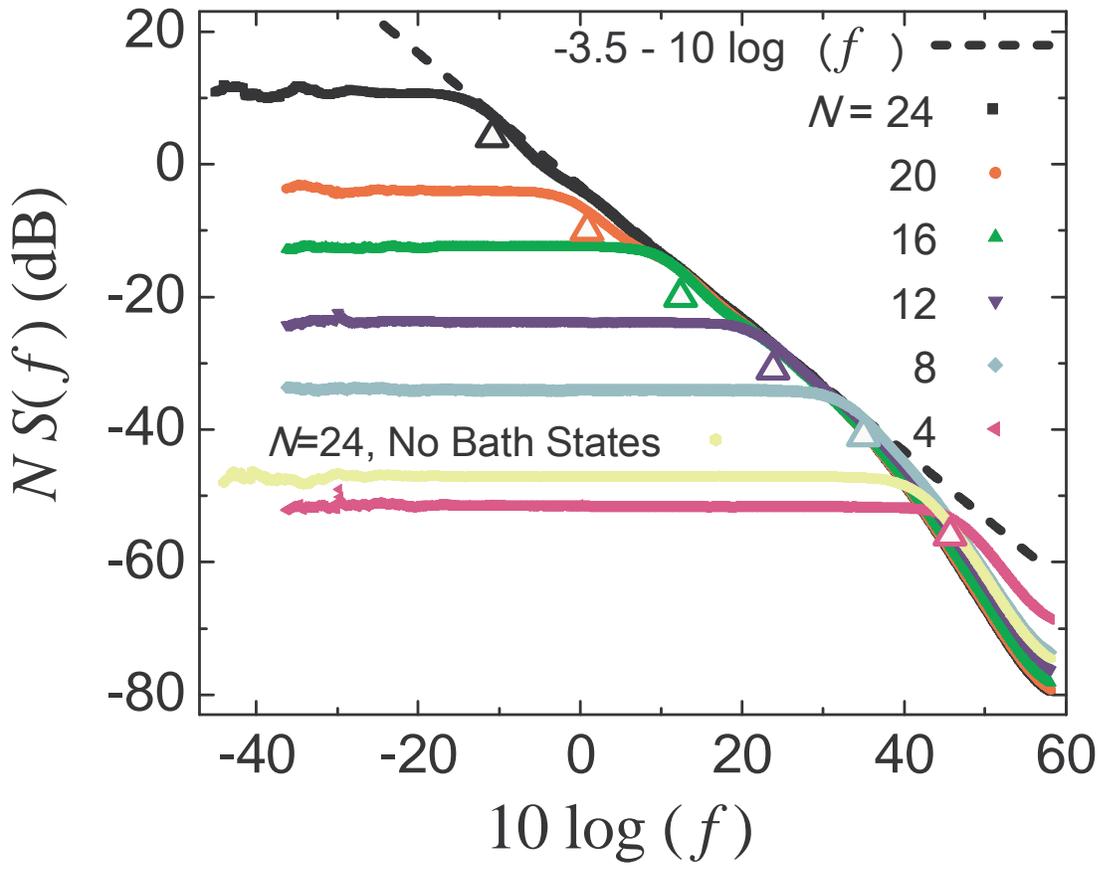

Fig. 3

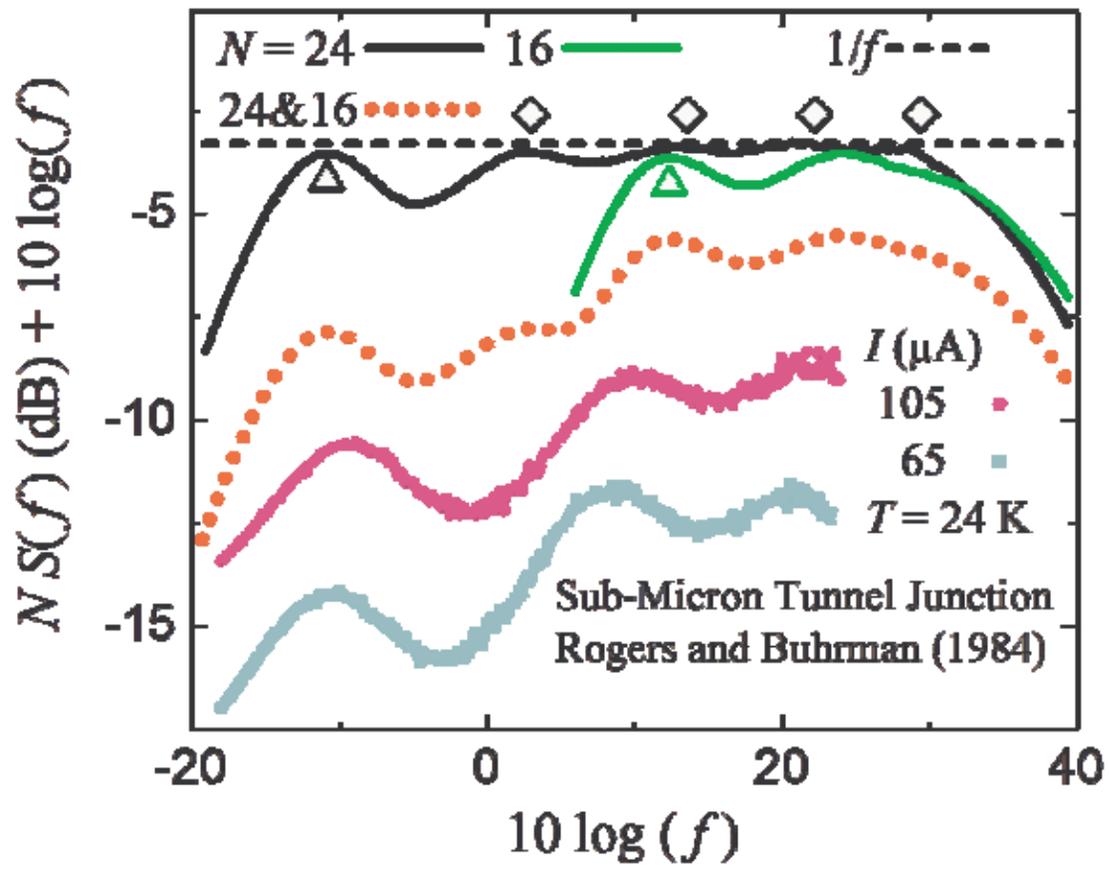

**Fig. 4**